\documentclass[twocolumn,showpacs,preprintnumbers,amsmath,amssymb,prb, superscriptaddress,floatfix]{revtex4}

\usepackage{graphicx}
\usepackage{dcolumn}
\usepackage{bm}
\usepackage{amsmath, amssymb, graphics, setspace}
\usepackage{hyperref}
\usepackage{color}

\newcommand{\mathsym}[1]{{}}
\newcommand{\unicode}[1]{{}}
\newcommand{\be}{\begin{eqnarray}}
\newcommand{\ee}{\end{eqnarray}}

\begin{document}


\title{Surface magnetism in a chiral $d$-wave superconductor with hexagonal symmetry}

\author{Jun Goryo}
\affiliation{Department of Mathematics and Physics, Hirosaki University, 060-8561 Hirosaki, Japan }
\author{Yoshiki Imai}
\affiliation{Department of Applied Physics, Okayama University of Science, 1-1 Ridaicho, Kita-Ku, 700-0005 Okayama, Japan}
\author{W. B. Rui}
\affiliation{Max-Planck-Institute f\"{u}r Festk\"{o}rperforschung, Heisenbergstrasse 1, D-70569 Stuttgart, Germany}
\author{Manfred Sigrist}
\affiliation{Institute f\"{u}r Theoretische Physik, ETH Z\"{u}rich, 8093 Z\"{u}rich, Switzerland}
\author{Andreas P. Schnyder}
\affiliation{Max-Planck-Institute f\"{u}r Festk\"{o}rperforschung, Heisenbergstrasse 1, D-70569 Stuttgart, Germany}

\date{\today}

\begin{abstract}


Surface properties are examined in a chiral $d$-wave superconductor with hexagonal symmetry, 
whose one-body Hamiltonian possesses intrinsic spin-orbit coupling identical to the one characterizing  
the topological nature of the Kane-Mele honeycomb insulator. In the normal state, spin-orbit coupling gives rise to spontaneous surface spin currents,   
 whereas  in the superconducting state, besides the spin currents,  there exist 
 also charge surface currents, due to chiral pairing symmetry. 
Interestingly, the combination of these two currents results in a surface spin polarization, 
whose spatial dependence is markedly different on the zigzag and armchair surfaces.
We discuss various potential candidate materials, such as SrPtAs, which may exhibit these surface properties.
 
\end{abstract}

\maketitle

\emph{Introduction.} Chiral superconductivity is becoming an increasingly hot topic 
in condensed matter physics. Chiral superconducting states break
time-reversal symmetry spontaneously and support
chiral surface states, which are of a topological origin. 
Therefore, they can be viewed as  superconducting   
analogs of the quantum Hall state  ~\cite{Volovik-text}. 
The spin-triplet chiral $p$-wave ($p_x \pm i p_y$-wave) state is experimentally observed in the $A$ phase of superfluid $^3$He thin films ~\cite{Yamashita-etal}.  
It is also the most plausible candidate for pairing symmetry 
in Sr$_2$RuO$_4$ ~\cite{maeno_review_JPSJ_12}. 
Its realization in the $\nu=5/2$ state of the fractional quantum Hall effect 
is of great interest due to quasiparticle (vortex) 
excitations obeying non-abelian statistics. ~\cite{Read-Green,5/2FQHE-exp1,5/2FQHE-exp2}.
Another chiral superconducting state is the spin-singlet chiral $d$-wave ($d_{x^2-y^2} \pm i d_{xy}$-wave) state.
Although this state has not yet been observed experimentally, there are many potential candidate materials, such as 
heavily doped graphene ~\cite{Black-Schaffer-Honerkamp},  water-intercalated sodium cobaltates ~\cite{Kiesel-Platt-Hanke-Thomale}, 
and SrPtAs ~\cite{Nishikubo-Kudo-Nohara,Biswas-etal,Goryo-Fischer-Sigrist,Fischer-etal,Fischer-Goryo,remark}.   
All of these materials exhibit hexagonal symmetry, which results 
in the degeneracy of $d_{x^2-y^2}$- and $d_{xy}$-wave pairing components and plays an important role in the stability of the chiral $d$-wave state. 

The aim of this paper is to study surface magnetism in chiral $d$-wave superconductors with hexagonal symmetry. 
In particular, we are interested in the interplay among surface magnetism, surface spin and charge currents,
and the topology of the chiral $d$-wave pairing state.
To study this, we 
 consider a generic tight-binding model with hexagonal symmetry
and spin-orbit coupling compatible with the crystal lattice symmetries. 
Similar to  the Kane-Mele (KM) topological insulator ~\cite{Kane-Mele},
the spin-orbit coupling in this model leads to spontaneous surface spin currents
in the normal state.
Since these spin currents are carried by  states well below the Fermi level, they persist
in the superconducting state. 
To study the surface properties of the superconducting state, we employ a self-consistent Bogoliubov-de Gennes (BdG) 
approach for   slab-shaped systems with zigzag and armchair surfaces.
We compute the charge currents ~\cite{chiral-current,Kalline,Tada-Nie-Oshikawa}  carried by the surface states of
the chiral $d$-wave pairing state. Interestingly, we find that the coexistence of spin and charge currents causes a spin polarization 
spontaneously at the sample surface, similar to the chiral $p$-wave superconductor Sr$_2$RuO$_4$ ~\cite{Imai-Wakabayashi-Sigrist}. 
The spatial dependence of the surface spin polarization shows a significant difference in
zigzag and armchair surfaces. 
We note that the superconducting state of the considered hexagonal lattice model belongs to class $D$ of the 
topological classification ~\cite{Schnyder-etal}, just as the chiral $p$-wave superconductor of Ref. ~\cite{Imai-Wakabayashi-Sigrist}.  
We expect that our findings are applicable to any class $D$ superconductor with spin-orbit coupling of the KM-model type. 

\begin{figure}
\begin{center}
\centering
\includegraphics[width=\linewidth]{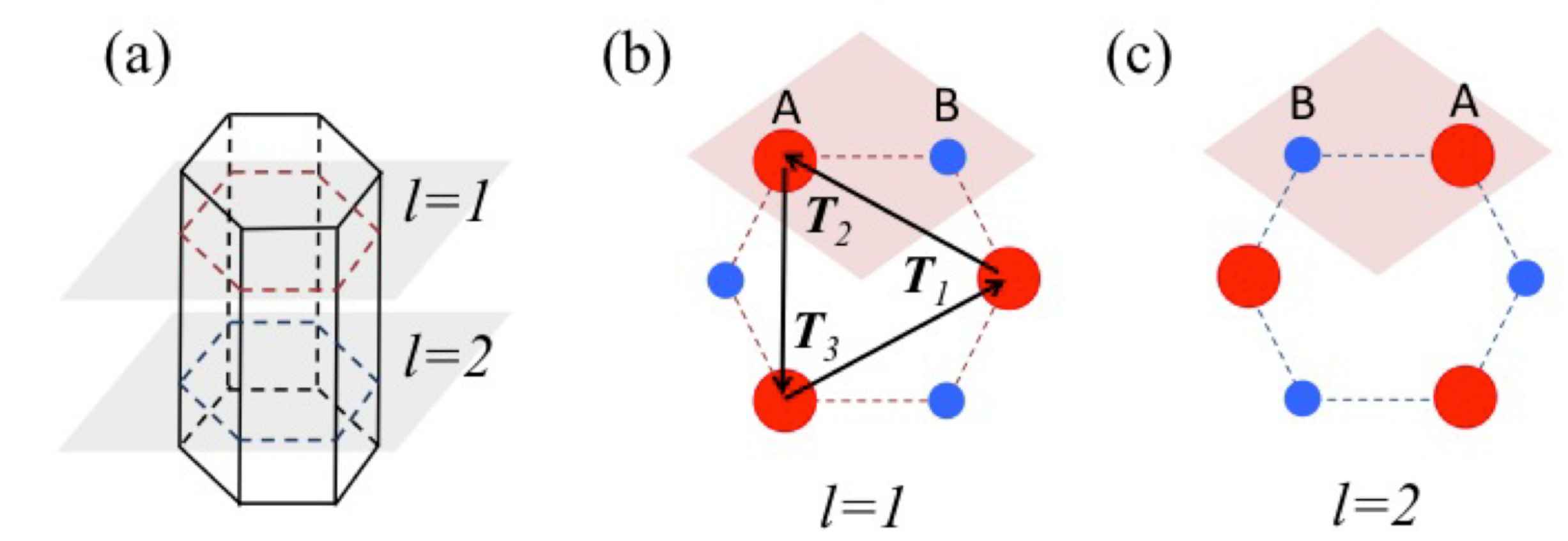}
\end{center} 
\caption{\label{mFig1}
Bilayer honeycomb lattice. (a) Whole unit cell with two distinct honeycomb sublayers.
Top view of the two honeycomb sublayers with (b) $l=1$, and (c) $l=2$.}
\label{hexagon}
\end{figure}

\emph{Model definition.}
Motivated by the hexagonal pnictide superconductor SrPtAs ~\cite{Nishikubo-Kudo-Nohara},
we consider a  double-layer hexagonal crystal structure with two honeycomb sublayers depicted in Fig.~\ref{hexagon}.
The two sublattices A and B of each honeycomb layer are taken to be inequivalent, 
such that each sublayer is noncentrosymmetric (i.e., locally noncentrosymmetric). However, 
the entire structure   has a global inversion center, since the sublattices A and B are exchanged
in the two sublayers. To capture the essential physics of such a bilayer honeycomb lattice,
we consider for simplicity only the electron hopping  on sublattice A (for instance, Pt sites in SrPtAs),
while sublattice B only plays the implicit role of breaking local inversion symmetry. 
The electrons in this system have therefore two internal degrees of freedom: spin and sublayer indices, $\sigma=\uparrow, \downarrow$ 
and $l=1,2$, respectively. With these assumptions, the tight-binding model for the normal state is described by $H_0=H_t+H_c+H_{c2}+H_{\lambda}$,
with
\begin{subequations}  \label{normal_state_Ham}
\be
H_t&=&t\sum_{n \delta l \sigma} \left(c_{\bm r_n l\sigma}^\dagger c_{\bm r_n + \bm T_\delta l\sigma}+c.c.\right)
-\mu\sum_{n l \sigma} c_{\bm r_n l\sigma}^\dagger c_{\bm r_n l\sigma},
\nonumber\\
H_c&=&t_c \sum_{n\sigma} (c_{\bm r_n 1 \sigma}^\dagger+c_{\bm r_n-\bm c 1 \sigma}^\dagger )
\big(c_{\bm r_n 2 \sigma}+c_{\bm r_n-\bm T_2 2 \sigma}
\nonumber\\
&& \qquad \qquad
+c_{\bm r_n+\bm T_3 2 \sigma}\big) 
 +c.c. ,
\nonumber\\
H_{c2}&=&t_{c2}\sum_{n l \sigma} c_{\bm r_n l\sigma}^\dagger c_{\bm r_n + \bm {c} l\sigma}+c.c.,
\ee
where $c_{\bm r_n l\sigma}^\dagger$ ($c_{\bm r_n l\sigma}$) stands
for the electron 
creation (annihilation) operator,
$\bm r_n$  indicates the position of the $n$th unit cell,
$\bm c$ is the lattice vector along the $c$ axis,
and $\bm T_1=(a,0,0)$, $\bm T_2=(-a/2,\sqrt{3}a/2,0)$, and $\bm T_3=(-a/2,-\sqrt{3}a/2,0)$ 
are the three nearest-neighbor in-plane bond vectors.
Besides the chemical potential energy $\mu$, Hamiltonian~\eqref{normal_state_Ham} contains 
three types of spin-independent hoppings: in-plane $t$, nearest-layer (intersublayer) $t_c$, 
and next-nearest-layer (intrasublayer) $t_{c2}$.   
In addition, due to the local lack of inversion center in each honeycomb layer, 
we have the locally antisymmetric spin-orbit coupling 
\be
H_{\lambda}&=&i \lambda\sum_{n \delta}\sum_{l \sigma} (-1)^l s_\sigma  c_{\bm r_n l\sigma}^\dagger c_{\bm r_n + \bm T_\delta l\sigma}+ c.c., 
\label{lasoc}
\ee
\end{subequations}
where $s_{\uparrow, \downarrow}=\pm 1/2$. 
Since this term is diagonal with respect to $S_z$, Hamiltonian~\eqref{normal_state_Ham} conserves $U(1)_z$ symmetry. 
However, this conservation is only approximate, as it is not protected by the crystalline symmetries. Indeed, 
there exist spin-dependent next-nearest-layer hopping terms which break $U(1)_z$ symmetry ~\cite{Fischer-Goryo}.   
But these higher-order terms are expected to be small, which
is confirmed by band-structure calculations for the case of SrPtAs ~\cite{Youn-etal}.  
For our numerical calculations we use parameters 
appropriate for the dominant band of SrPtAs, which consists of
two spin-orbit-split Fermi sheets (an open cylindrical one and a closed cigar-shaped one) centered around the Brillouin zone corners ~\cite{Youn-etal}.

In passing we note that Hamiltonian~\eqref{normal_state_Ham} also describes the normal state of
UPt$_3$ ~\cite{Shiozaki-Yanase}.
Furthermore, we observe that the spin-orbit coupling (\ref{lasoc}) is equivalent 
to the intrinsic spin-orbit coupling of the KM honeycomb quantum spin Hall insulator ~\cite{Kane-Mele}.  
That is, by identifying the sublayer index in Eq.~\eqref{normal_state_Ham} with the sublattice degree of freedom of the KM model,
the latter maps onto Hamiltonian~\eqref{normal_state_Ham} in the limit of letting $t, t_{c2}, \mu, c \rightarrow 0$.  
For this reason, our model~\eqref{normal_state_Ham}  describes also the physics of graphenelike structures 
with large spin-orbit coupling, such as stanene ~\cite{stanene_nat_mat_15} or  graphene on   WS$_2$ ~\cite{WS2_nat_commun}. 
 
From the analogy with the KM model ~\cite{Kane-Mele}, it follows that Hamiltonian~\eqref{normal_state_Ham} exhibits 
a large spin Hall conductivity $\sigma_{xy}^s$. Indeed, using the Kubo formula
and parameter values appropriate for SrPtAs ~\cite{Youn-etal},  
we find  that  
$  \sigma_{xy}^s \simeq -120 \hbar / (e \Omega\textrm{cm})$ [see Appendix A]. 
This value of $\sigma_{xy}^s$ is comparable to that of Pt, which is a typical spin Hall metal ~\cite{Kimura-etal}.  
We remark that a nonzero $z$ component of spin-orbit coupling
is important for   a large spin Hall conductivity. 
A locally noncentrosymmetric system with, for example, only   
staggered Rashba spin-orbit coupling ~\cite{Maruyama-etal} does not
have a net spin Hall conductivity, since the contributions from the two  layers cancel out.

\begin{figure}
\begin{center}
\includegraphics[width=\linewidth]{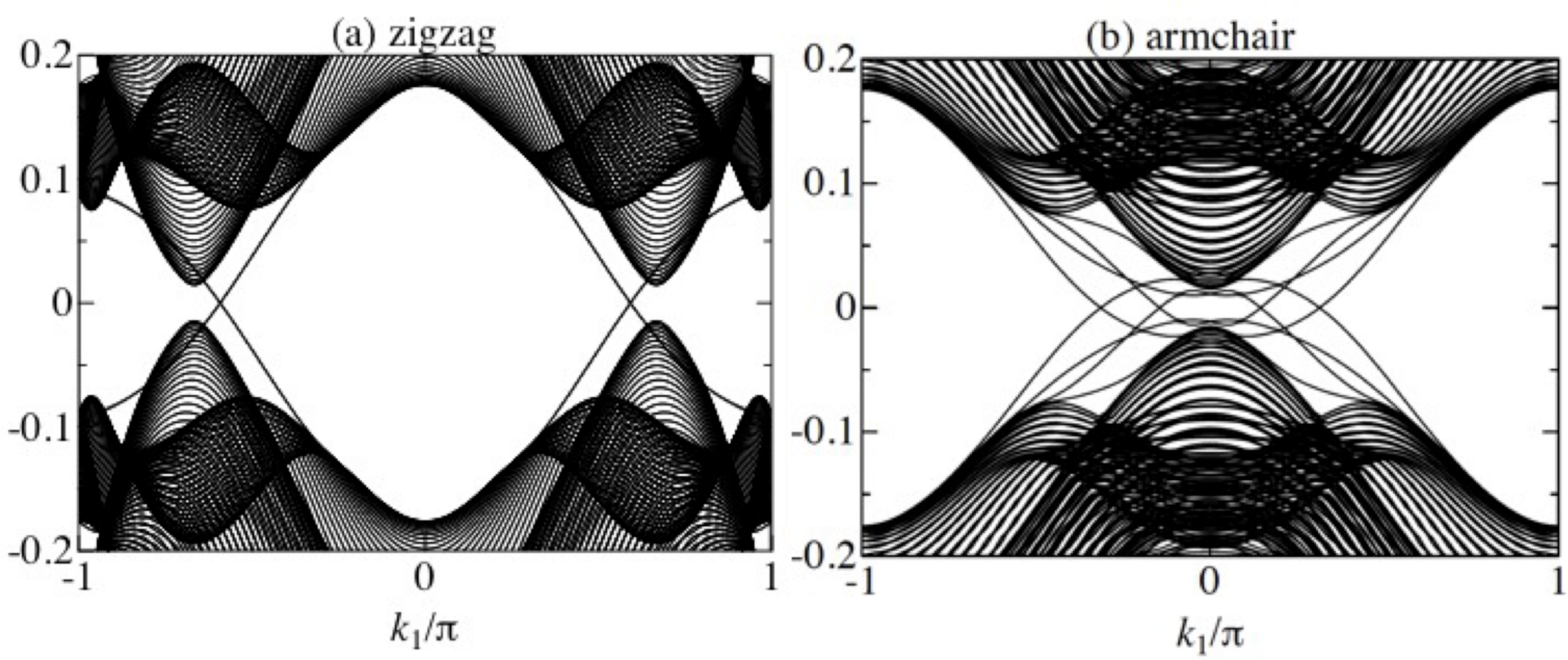}
\end{center}
\caption{Surface spectrum at $k_z=0$ of the chiral $d$-wave superconductor 
for (a) the zigzag and (b) the armchair slab.
}
\label{BdGspectrum}
\end{figure}

To introduce unconventional spin-singlet superconductivity in our model system,
we add to Hamiltonian~\eqref{normal_state_Ham}
a density-density-type pairing interaction between 
two electrons on in-plane nearest-neighbor sites in each honeycomb sublayer,  i.e.,
\be \label{interactionTerm}
H_{\textrm{int}} = \frac{U}{2} \sum_{n\delta}\sum_{l\sigma\sigma'} n_{\bm r_n l \sigma} n_{\bm r_n + \bm T_\delta l\sigma'},
\ee
with   $U <0$ and $n_{\bm r_n l\sigma}=c_{\bm r_n l\sigma}^\dagger c_{\bm r_n l\sigma}$. 
The attractive pairing interaction~\eqref{interactionTerm} is decoupled as usual by  BCS-type mean fields  with 
the gap functions
$
\Delta_{\bm r_n l}^{(\delta)}=U / 2 \left<c_{\bm r_n+\bm T_\delta l \downarrow} c_{\bm r_n l \uparrow}
- c_{\bm r_n+\bm T_\delta l \uparrow} c_{\bm r_n l \downarrow} \right>,
$
which correspond to in-plane spin-singlet pairing in the three different directions ${\bm T}_\delta$. 
To determine the order parameters $\Delta_{\bm r_n l}^{(\delta)}$ we
numerically solve the self-consistent gap equation  with $U =-0.5$. 
(We also considered smaller values of $U$, e.g., $U=-0.2$, which leads to qualitatively similar results.)
We find that the stable homogeneous solutions of the gap equation satisfy
$
\Delta^{(1)}=\omega^2\Delta^{(2)}=\omega \Delta^{(3)}=\Delta
$, where $\omega=e^{i 2 \pi /3}$. 
This corresponds to chiral $d$-wave pairing, since 
 the state has   eigenvalue $e^{4 \pi i}$ with respect to $2\pi$ rotations about the $z$ crystal axis.
This chiral $d$-wave state has a nontrivial topology, which is characterized
by a quantized Chern number defined in terms of an integral
along a two-dimensional submanifold of the three-dimensional Brillouin zone. 
Choosing, for example, $k_z=0$ as the integration plane, the Chern number 
evaluates to four  ~\cite{Fischer-etal}. By the bulk-boundary correspondence, this
indicates that four chiral surface states appear 
at boundaries that are parallel to the (001) direction (see Fig.~\ref{BdGspectrum}).

\begin{figure}
\begin{center}
\includegraphics[width=\linewidth]{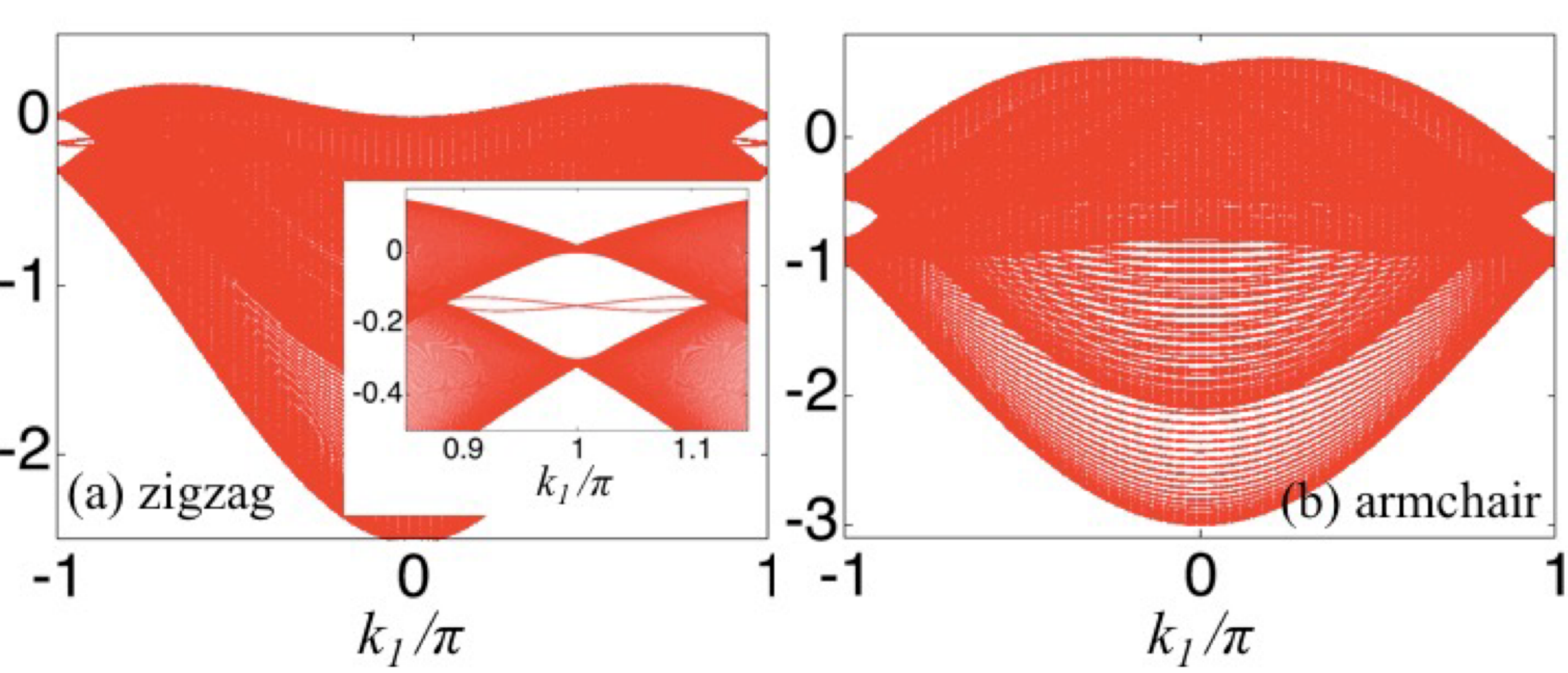}
\end{center}
\caption{Normal state spectrum at $k_z=0$ for (a) the zigzag  and (b) the armchair slab. We clearly see the helical edge spectrum in the zigzag case.}
\label{NormalSpectrum}
\end{figure}

\emph{Surface properties of the normal state.}
Before studying the surface properties of the superconducting state, let us first analyze
 the surface spectrum of the normal state (i.e., $U=0$) at zero temperature.
For that purpose, we numerically diagonalize the normal state Hamiltonian~\eqref{normal_state_Ham} in a slab geometry 
of dimensions $M \times N \times N_z = 100 \times 100 \times 100$
with zigzag and armchair surfaces. 
The zigzag (armchair) surfaces are implemented by imposing periodic boundary conditions in the $z$ direction and the
$\bm T'_1$ direction, while imposing open boundary conditions along 
the  $\bm T'_2$ direction [the definitions of $\bm T'_1$ and $\bm T'_2$, and the Fourier transform of the Hamiltonian \eqref{normal_state_Ham} can be found in Appendix B].
We note that for this slab geometry there are $2N =200$ bands for each spin sector.
In Fig.~\ref{NormalSpectrum} we present the band structure for the up spin sector at $k_z=0$.
We clearly see in the zigzag case that the $N$ and $(N+1)$th energy bands show a level crossing at $k_1=\pi$, 
which is the trace of the helical edge mode of the KM model. We find that the $N$th band carries a chiral flow of up spin electrons 
at the front surface, while the $(N+1)$th band carries the counterflow at the back surface. 
Although it is less clear in Fig. \ref{NormalSpectrum}, we also have a level crossing hidden inside the bulk bands in the armchair case.  

\begin{figure}
\begin{center}
\includegraphics[width=\linewidth]{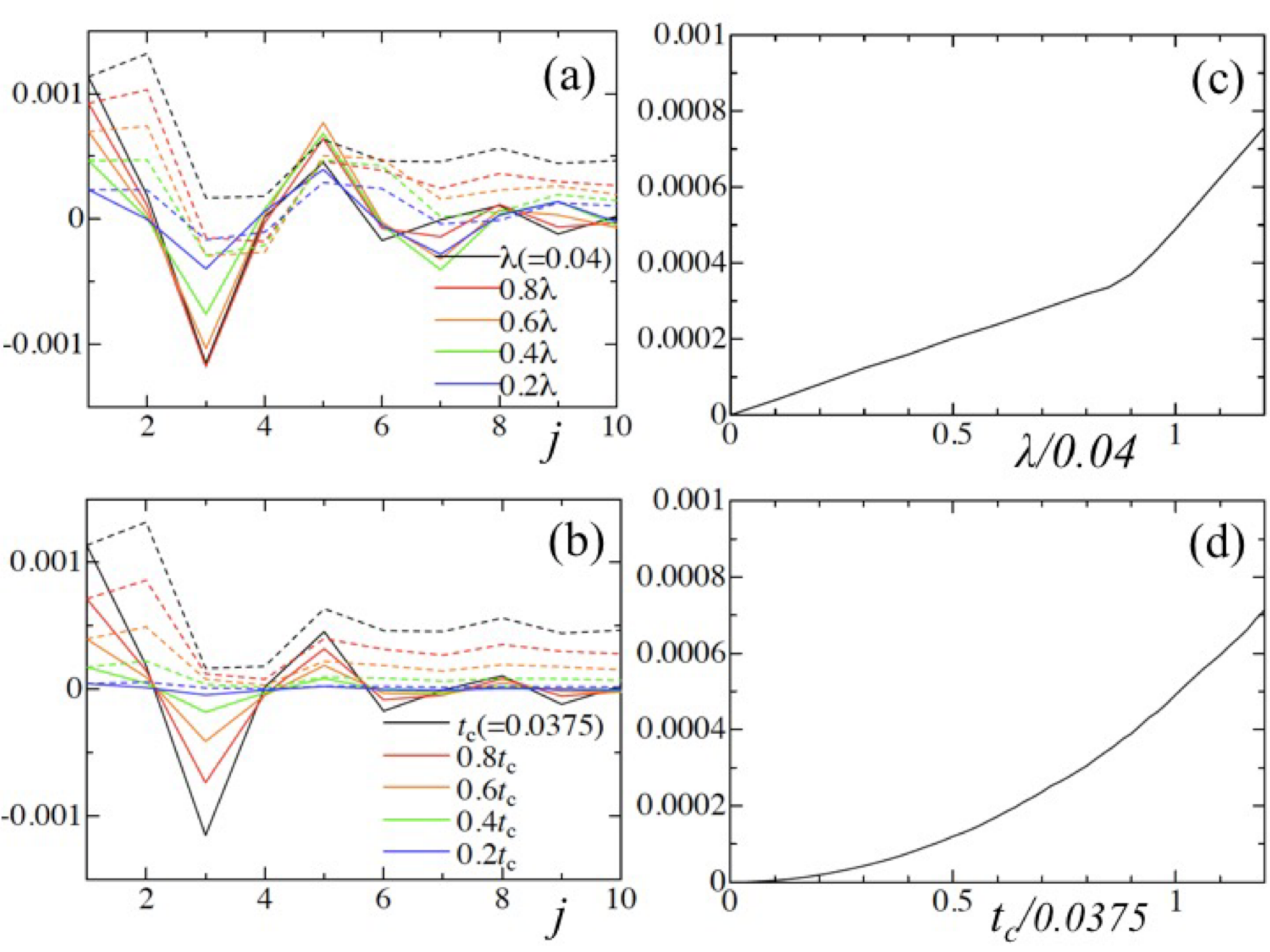}
\end{center}
\caption{Spatial dependence of the spin current $\langle J^s_j \rangle$ 
at the zigzag surface in units of $ta/2$ 
for several values of (a) spin-orbit coupling $\lambda$ and (b) nearest-inter-sublayer hopping  $t_c$ with $T=U=0$.
Dotted lines are 
integrated values of the solid lines from $j=0$. 
 Panels (c) and (d) show the $\lambda$- and $t_c$-dependence of $J^s_{\rm sum}\equiv \sum_{j=1}^{N/2}\langle J^s_j \rangle$. 
$\lambda=0.04$ and $t_c=0.0375$ are the suggested values for the dominant band in SrPtAs ~\cite{Youn-etal}.  
For the armchair surface the results are qualitatively similar. }
\label{Normal-Js}
\end{figure}

In order to quantify the surface spin current, 
we compute the thermal expectation value of the spin dependent velocity operator
$\langle v_{jl \sigma} \rangle$  at the $j$th site ($j=1, 2, ... ,N$) in the $l$th layer.
Using Eq.~\eqref{normal_state_Ham}, we find
  \be
\langle v_{jl \sigma} \rangle
=\frac{1}{MN_z}\sum_{\bm kj'l'} \left\langle c_{\bm k jl\sigma}^\dagger \left(\frac{\partial \hat{h}_{0 \bm k \sigma}}{\partial k_1}\right)_{jl;j'l'} c_{\bm k j'l'\sigma} \right\rangle, 
\ee
where $\hat{h}_{0 \bm k \sigma}$ with $\bm k=(k_1,k_z)$ is the $2N\times 2N$ one-body Hamiltonian matrix with spin $\sigma$.  
With this, the charge and spin currents are defined as 
$\langle J_{j}^c \rangle=-e \sum_{l\sigma} \langle v_{jl\sigma} \rangle$ and 
$\langle J_{j}^{s} \rangle=\sum_{l\sigma} s_\sigma \langle v_{jl\sigma} \rangle$. 
Figure \ref{Normal-Js} shows the spatial distribution of $\langle J_{j}^{s} \rangle$. 
(Due to time-reversal symmetry there is no charge current in the normal state.)
We also plot in Fig.~\ref{Normal-Js} the $\lambda$ and $t_c$ dependence of the spin current 
and find that a nonzero value of both is crucial for a finite spin current. 
In fact, $J_{\rm sum}^s$ is to leading order proportional to $\lambda$ and $t_c^2$ [Figs.~\ref{Normal-Js}(c) and~\ref{Normal-Js}(d)], 
which is in agreement with the analytical expression  of $\sigma_{xy}^s$ [see Eq.~(A5) in Appendix A].  

\begin{figure}
\begin{center}
\includegraphics[width=\linewidth]{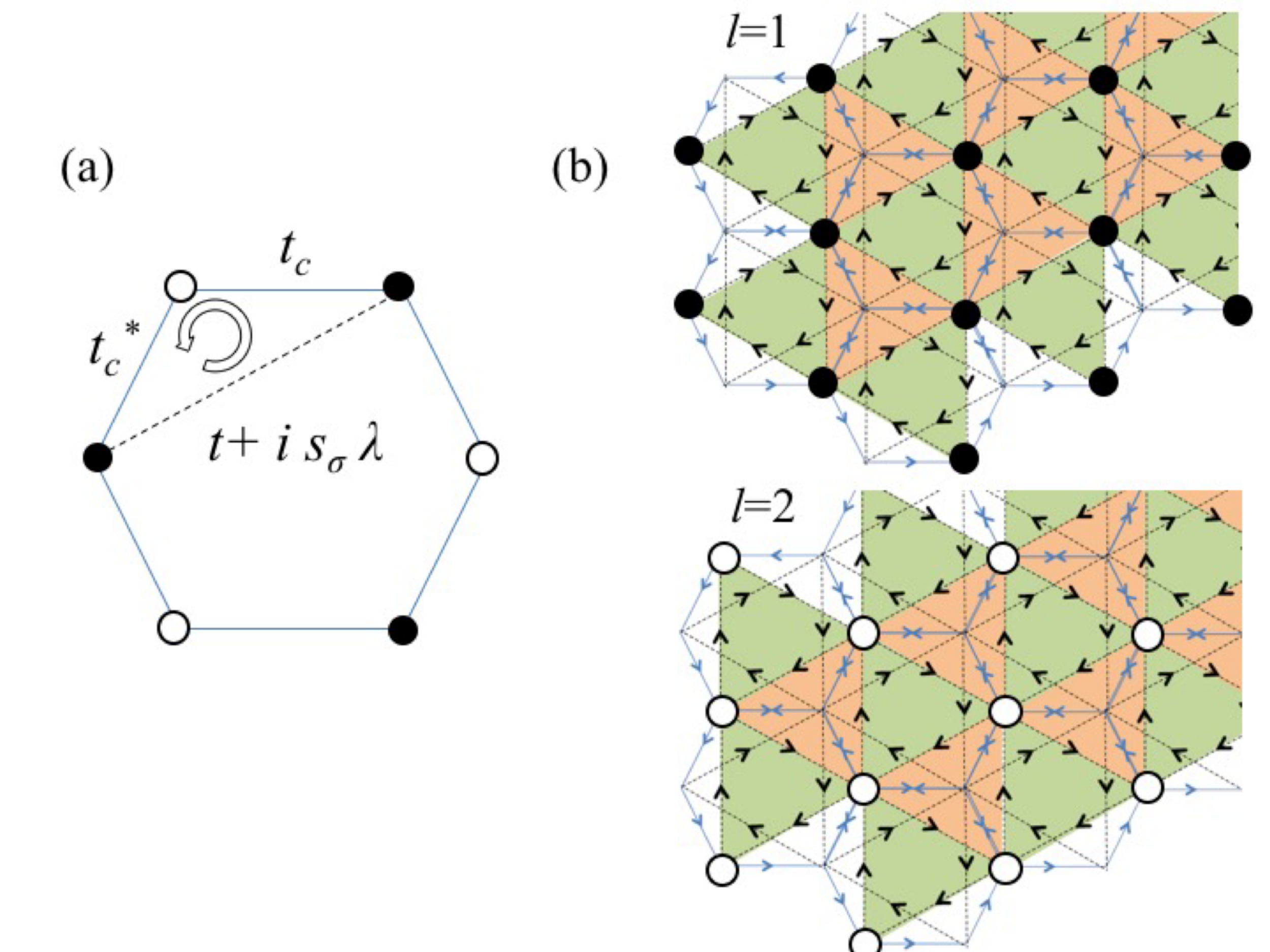}
\end{center}
\caption{
 Panel (a) shows the elemental 
 loop for acquiring the spin-dependent phase $\phi_\sigma$.
 The black dots (open circles) on the corners of the hexagon indicate   the electron hopping sites in the $l=1$ ($l=2$) sublayer. 
The arrows in panel (b) show the spin current distribution for the case 
 $\phi_\uparrow = - \phi_\downarrow >0$, 
and the colors of triangles show the circulation direction of the spin currents on the intrasublayer bonds (green: clockwise; orange: anti-clockwise).
}
\label{spin-dependent-AB}
\end{figure}

Analogous to the model of Sr$_2$RuO$_4$ ~\cite{Imai-Wakabayashi-Sigrist}, the origin of 
this surface spin current can be understood as follows. We see from Fig.~\ref{spin-dependent-AB}(a) that an electron with spin $\sigma$ acquires a phase 
$\phi_\sigma= \tan^{-1} s_\sigma\lambda/t$ when it travels in the anticlockwise direction along a triangle enclosed by two interplane nearest-neighbor bonds (edges of the hexagon) 
and an in-plane nearest-neighbor one (dotted line). This results in a circular spin current along the triangle. 
In the whole system, the spin currents on the inter-plane bonds cancel out in the bulk, but survive at the sample surface. 
The currents on the in-plane bonds (black arrows in Fig. \ref{spin-dependent-AB}), on the other hand, are not canceled out even in the entire bulk. 
However, their circulation direction, depicted by colored triangles in Fig. \ref{spin-dependent-AB}, shows a staggered pattern in each sublayer,  
and hence they are smeared out in the long-wavelength limit.
This is anticipated since our model corresponds to the metallic phase of  
the KM model, which is the spin analog of the Haldane model for the quantum Hall effect without net magnetic flux. 
Using the above considerations, we can define in the long-wavelength limit a
spin-dependent current density ~\cite{Imai-Wakabayashi-Sigrist} 
$\bm j_{\sigma}(\bm r)=\bm \nabla \times \hat{\bm z} \Phi_{\sigma}(\bm r)$,
where $\Phi_{\sigma}(\bm r)=-\Phi_{-\sigma}(\bm r)$ is the ``spin-dependent" flux associated with $\phi_\sigma$. 
It follows that there exists a finite spin current density 
 $\bm j_\uparrow - \bm j_\downarrow$ 
whenever $\Phi_{\sigma}(\bm r)$ is not uniform, 
which is the case at the surface.

\begin{figure}
\begin{center}
\includegraphics[width=\linewidth]{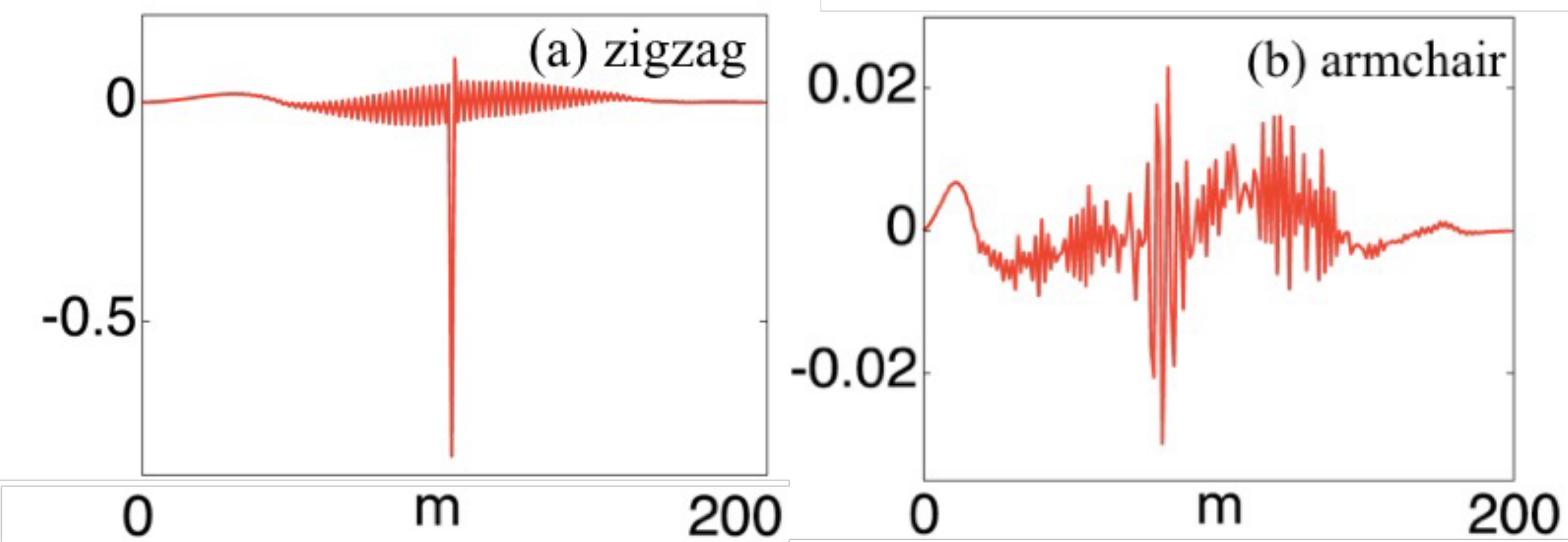}
\end{center}
\caption{Contribution to the surface spin current $\langle J^s_{j=1} \rangle$ in units of $ta/2$ from the $m$th Bloch state at $T=U=0$ for 
(a) the zigzag and (b) the armchair surface. 
}
\label{mode-vs-Js}
\end{figure}

From Fig.~\ref{mode-vs-Js} we can see that the surface spin current is carried by Bloch bands well below 
the Fermi level. The surface current therefore persists in the superconducting phase, since  superconductivity does not affect 
 states far from the Fermi level. For the zigzag surface, there is a sharp peak at the $N$th energy band corresponding to 
a counterpart of the helical edge states. 

\emph{Surface properties of the superconducting state.}
Next, we study the surface properties of the superconducting state. 
For that purpose, we numerically solve the self-consistent BdG equations for
a slab geometry with zigzag and armchair surfaces.
We find that the chiral $d$-wave state is a stable solution also in slab geometries.
Since the BdG Hamiltonian possesses a mirror symmetry which lets $k_z \to - k_z$ and exchanges 
the two sublayers, each eigenstate at $k_z=0$ can be labeled by a mirror eigenvalue.
This mirror symmetry has two eigensectors at $k_z=0$ with two
chiral surface states in each eigensector (Fig.~\ref{BdGspectrum}).
We find that the two mirror eigensectors of the BdG Hamiltonian transform into each other under particle-hole symmetry,
since the gap function is even under the mirror reflection $k_z \rightarrow -k_z$ ~\cite{Sato-Ando}.  
That is, while the entire Hamiltonian is particle-hole symmetric, each mirror eigensector 
breaks this symmetry.
 Therefore, the energy spectra of the two mirror eigensectors may split. Indeed, 
 this is observed in Fig. \ref{BdGspectrum}, which shows the energy spectra at $k_z=0$. 
For the armchair surface the splitting is large enough to see four chiral edge states;
for the zigzag surface the splitting is also nonzero, but too small to be seen
in Fig.~\ref{BdGspectrum}(a).

\begin{figure}
\begin{center}
\includegraphics[width=\linewidth]{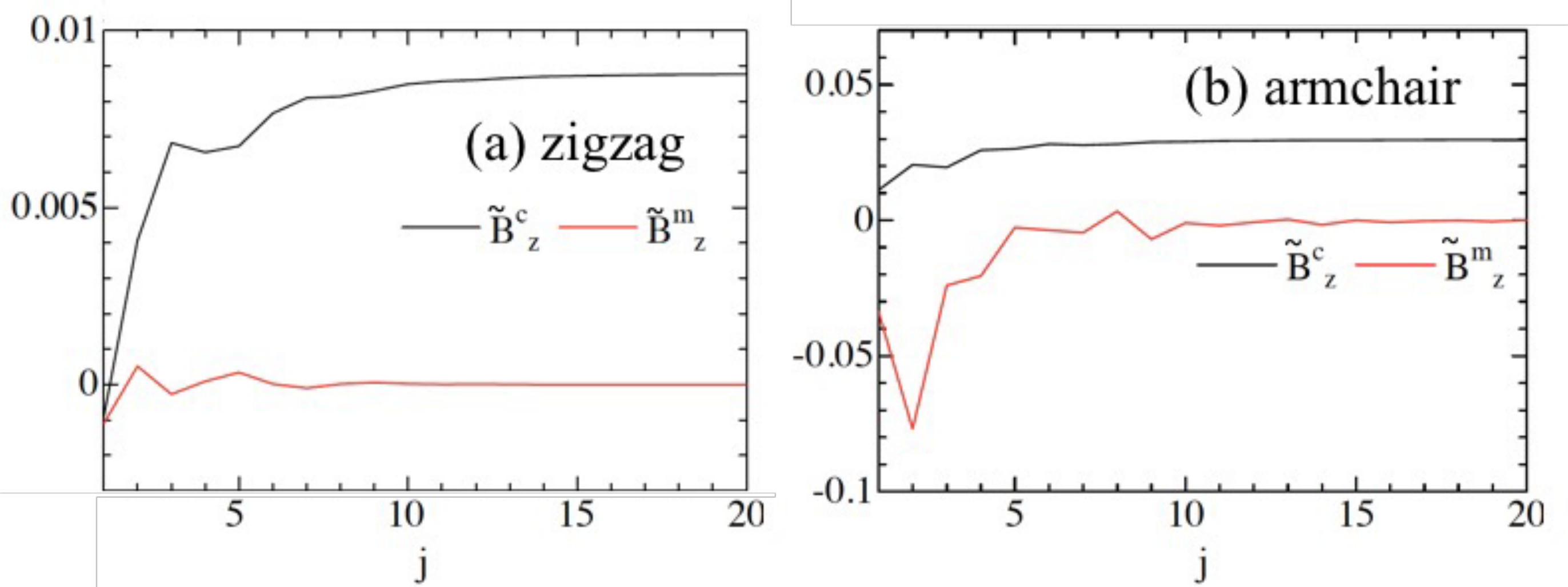}
\end{center}
\caption{
Dimensionless magnetic fields at (a) the zigzag and (b) the armchair surface of the
chiral $d$-wave superconductor at $T=0$. 
The black lines represent the magnetic field $\tilde{B}_{zj}^c$ induced by the chiral surface current,
while the red lines represent the magnetic field $\tilde{B}_{zj}^m$ produced by the spin polarization.
Note that the Meissner effect is neglected.}
\label{Bc-Bm}
\end{figure}

The chiral surface states of Fig.~\ref{BdGspectrum} carry a charge current ~\cite{chiral-current}.  
In addition,   the superconducting state  exhibits a surface spin current due to
states well below the Fermi level, which are not affected by superconductivity.
The coexistence of spin and charge 
currents leads to a spin polarization  at the sample surface. 
Both the charge current and the spin polarization produce  net magnetic fields, which are given by
 ~\cite{Imai-Wakabayashi-Sigrist} 
\begin{subequations} \label{ind-B}
\begin{eqnarray} \label{B_due_to_charge_c}
B_{zj}^c &=&\mu_0 \sum_{j'}^j \langle J^c_{j'}\rangle=\frac{e \mu_0 t a}{\hbar S_0}\tilde{B}_{zj}^c, 
\\
B_{zj}^m &=& -\frac{\mu_0}{\Omega_c} \sum_{l \sigma} g \mu_B s_\sigma n_{j \sigma}=\frac{\mu_0 \mu_B}{\Omega_c}\tilde{B}_{zj}^m, 
\end{eqnarray}
\end{subequations}
where $n_{j \sigma}=\sum_{\bm k l}\langle c_{\bm k j l \sigma}^\dagger c_{\bm k j l \sigma}\rangle/(MN_z)$, $\mu_0$ is the magnetic permeability,
and $\tilde{B}_{zj}^c$ and $\tilde{B}_{zj}^m$ are the dimensionless fields on layer $j$. 
The constants
 $S_0=ac$
and $\Omega_c=k a^2c$ ($k=1$ for zigzag and $k=\sqrt{3}$ for armchair cases) are the cross section and volume of the unit cell in the slabs, respectively.
Note that the Meissner screening is neglected here, which would introduce counter-propagating currents
on a length scale of the London penetration depth. 
In Fig.~\ref{Bc-Bm} the two magnetic fields are shown for the zigzag and armchair surfaces.
To compare the two fields, the dimensional prefactors $ e \mu_0  t a/\hbar S_0$ and $\mu_0 \mu_B/\Omega_c$ must be taken into account, 
which, however, turn out to be of the same order ~\cite{Imai-Wakabayashi-Sigrist}. 
Hence, we conclude that on the armchair surface the two fields tend to cancel each other, 
since they have a similar magnitude but opposite sign. 
This cancellation is independent of the chirality of the $d$-wave state,
since the direction of the spin polarization 
is coupled to the chirality through  spin-orbit coupling. 
On zigzag surfaces, on the other hand, there is no cancellation between
the two fields, since  $B_{zj}^m$ oscillates around zero.
 

\emph{Discussion.}
We have investigated  spin and charge currents at the surfaces of chiral $d$-wave superconductors with hexagonal symmetry
and intrinsic spin-orbit coupling. We have shown that the combination
of spin and charge currents leads to a spontaneous spin polarization at the surface of the superconductor.
Both the spin polarization and the chiral charge currents generate magnetic fields,
whose spatial dependence we have studied in detail.
These magnetic fields and spontaneous spin polarization could be observed experimentally using 
scanning Hall probe microscopy, 
electron spin resonance (ESR), or spin dependent tunneling. 

Our results are not only relevant for the hexagonal pnictide superconductor SrPtAs, but also for
doped graphenelike structures with strong spin-orbit coupling, such as stanene.
Recent muon spin rotation ($\mu$SR) experiments on polycrystalline SrPtAs ~\cite{Biswas-etal} have
detected spontaneous magnetic fields in the superconducting state,
which is qualitatively in agreement with our findings. 
Unfortunately, single crystals of SrPtAs are not yet large enough to allow
for  $\mu$SR measurements ~\cite{Nohara-private}.  
However, with these small single crystals it might be possible
to measure the predicted spin polarization using ESR or spin-dependent tunneling experiments.

The authors thank M. Nohara and M. H. Fischer for their useful discussions. This work was partially supported by JSPS KAKENHI Grant No.15H05885 (J-Physics).  
J.G. is grateful to the Pauli Center for Theoretical Physics of ETH Zurich for hospitality.

\newpage

\appendix

\section{The bulk spin Hall conductivity in the normal state}

As the system is analogous to the metallic phase of Kane-Mele (KM) model, 
we would have the spin Hall conductivity. In the Fourier space, 
\be
H_0&=&\sum_{\bm k,\sigma}\sum_{l,l'} c_{\bm k l \sigma}^\dagger (\hat{h}_{0\bm k \sigma})_{l l' } c_{\bm k l'\sigma}, 
\nonumber\\
\hat{h}_{0 \bm k \sigma}&=&\epsilon_{\bm k} \hat{\tau}_0+\bm g_{\bm k \sigma} \cdot \hat{\bm \tau},
\nonumber\\
\bm g_{\bm k \sigma}&=&({\rm Re}(t_{c\bm k}), -{\rm Im}(t_{c\bm k}), -2 s_\sigma \lambda_{\bm k}),
\nonumber\\
\epsilon_{\bm k}&=&-2 t \sum_{\delta=1}^3 \cos \bm k \cdot \bm T_\delta -2 t_{c2} \cos k_z c - \mu, 
\nonumber\\
t_{c \bm k}&=&t_c (1+e^{i \bm k \cdot \bm T_2}+e^{-i \bm k \cdot \bm T_3}) (1+e^{i k_z c})=t_{c-\bm k}^*,
\nonumber\\
\lambda_{\bm k}&=&\lambda \sum_{\delta=1}^3 \sin \bm k \cdot \bm T_\delta=-\lambda_{-\bm k},
\ee 
where $\hat{\tau}_0$ and $\hat{\bm \tau}$ are the $2\times2$ unit and Pauli matrices with sublayer indices $l$ and $l'$. 
The energy spectrum of $H_0$ is 
$\xi_{\bm k}^{\pm}=\epsilon_{\bm k}\pm \sqrt{|t_{c\bm k}|^2+\lambda_{\bm k}^2}$, and each branch $\pm$ has the Kramers degeneracy. 
The operators for charge and spin currents are
\be
\bm j&=&-e \sum_\sigma \bm v_\sigma,
\label{e-cur}
\\ 
\bm j_s&=&\sum_\sigma s_\sigma \bm v_\sigma, 
\label{s-cur}
\ee
where 
\be
\bm v_\sigma=\sum_{ll'} \left\{{\bm \nabla_{\bm k}}  \hat{h}_{0\sigma}\right\}_{ll'} c_{\bm k l \sigma}^\dagger  c_{\bm k l'\sigma}, 
\ee
is the spin-dependent velocity operator.  
We estimate the charge current-spin current correlation function and 
employ the Kubo formula for the spin Hall conductivity  $\sigma_{xy}^s$. We neglect the vertex correction 
and obtain the following form, which reminds us the Skyrmion number
\be
\sigma_{xy}^s&=&
\frac{e}{2\pi}\int \frac{d^3k}{8 \pi^2}\sum_{\sigma} s_\sigma \hat{\bm g}_{\bm k \sigma}\cdot \left(\partial_{k_x} \hat{\bm g}_{\bm k \sigma} \times \partial_{k_y} \hat{\bm g}_{\bm k \sigma}\right)\times
\nonumber\\
&&\left(f(\xi_{\bm k}^+)-f(\xi_{\bm k}^-)\right), 
\label{SHC}
\ee  
where $\hat{\bm g}_{\bm k \sigma}={\bm g}_{\bm k \sigma}/|{\bm g}_{\bm k \sigma}|$ and $f(\xi_{\bm k}^\pm)$ is 
the Fermi distribution function for each subenergy band. 
Note that $\sigma_{xy}^s$ is not quantized even at the zero temperature, 
because the $k$-integration is over the partially occupied states in the Brillouin zone (not the entire Brilloun zone). Using the tight-binding parameters 
for all the conduction bands in SrPtAs obtained from the band structure calculation ~\cite{Youn-etal}, we find $\sigma_{xy}^s \simeq -120 \hbar /(e\Omega {\rm cm})$ at $T=0$. 
This value is comparable to the spin Hall conductivity of Pt, which is a typical spin Hall metal ~\cite{Kimura-etal}.  

Eq. (\ref{SHC}) suggests that $\sigma_{xy}^s$ is proportional to $\lambda$ and $t_c^2$ when they are small. 
We should emphasize here that this behavior is consistent with the results in Fig. \ref{Normal-Js}. 

\section{Fourier transforms of the non-interacting Hamiltonian (1) in zigzag and armchair slabs}

\begin{figure}
\begin{center}
\includegraphics[width=9cm]{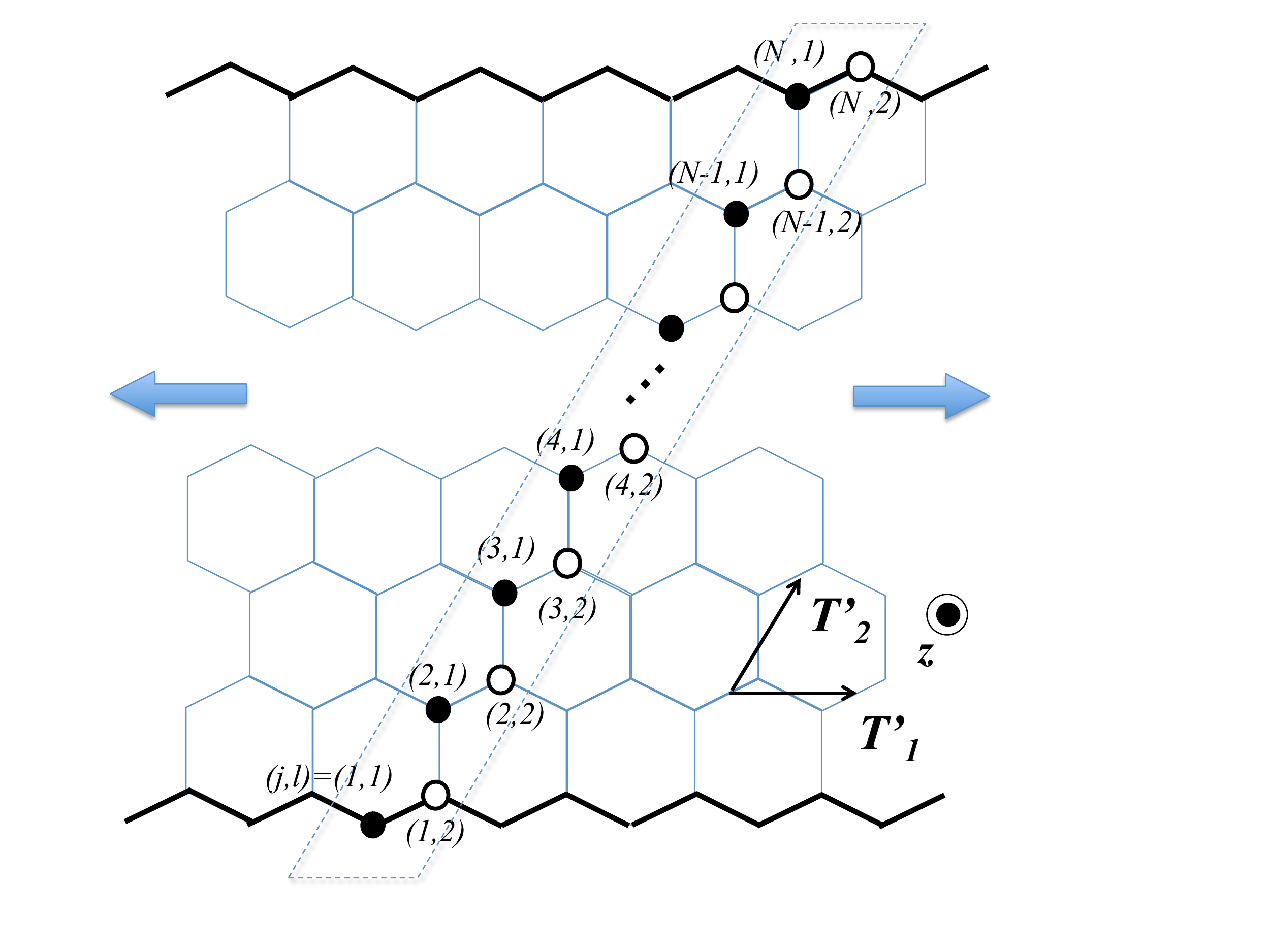}
\end{center}
\caption{The zigzag slab with lattice vectors $\bm T'_1=(1, 0, 0)$, $\bm T'_2=(1/2, \sqrt{3}/2, 0)$, and $\bm z=(0,0,1)$.}
\label{zigzag}
\end{figure}

\begin{figure}
\begin{center}
\includegraphics[width=9cm]{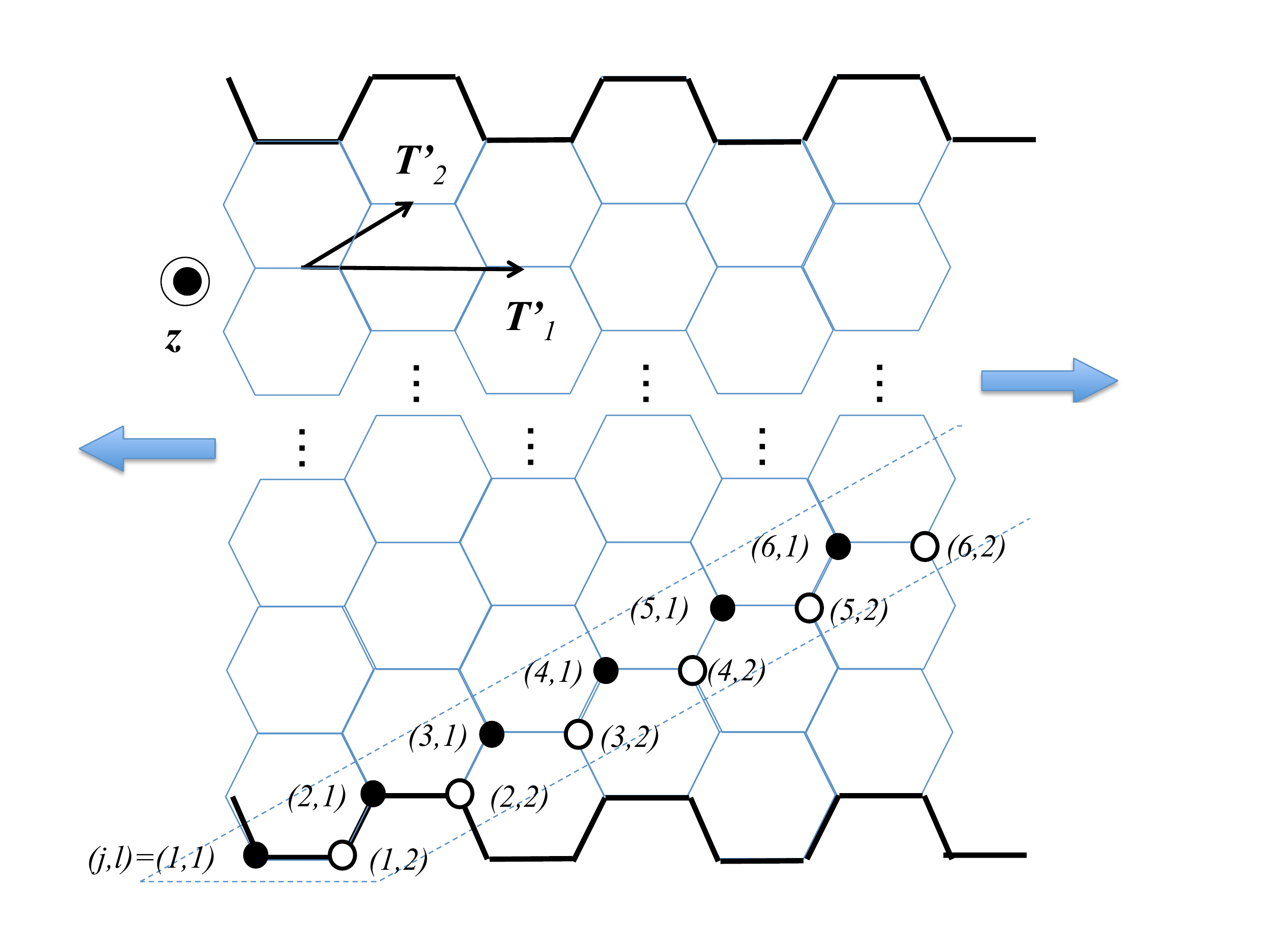}
\end{center}
\caption{The armchair slab with lattice vectors $\bm T'_1=(1, 0, 0)$ and $\bm T'_2=(1/2, 1/(2\sqrt{3}), 0)$, and $\bm z=(0,0,1)$.}
\label{armchair}
\end{figure}

 We show the Fourier transforms of the normal state Hamiltonian (1) in zigzag and armchair cases 
(see Figs. \ref{zigzag} and \ref{armchair}).  Let $c_{\bm k j l \sigma}^\dagger$ ($c_{\bm k j l \sigma}$ ) stands for the creation and annihilation operator of the electron at the $j$-th site in the $l$-th layer 
with momentum $\bm k=(k_1, k_z)$ and spin $\sigma$, and the results are as follows: 

\begin{itemize}

\item The zigzag case
\be
H_t&=& 2 t \sum_{\bm k j l \sigma} \left\{\cos k_1 c^\dagger_{\bm k j l \sigma} c_{\bm k j l \sigma}+ \right.
\nonumber\\
&&\left. \cos \frac{k_1}{2} \left(c^\dagger_{\bm k j l \sigma} c_{\bm k j+1 l \sigma} + c.c.\right) \right\}, 
\nonumber\\
H_c&=& t_c \sum_{\bm k j \sigma} \left( 2 \cos \frac{k_1}{2} e^{i \frac{k_z}{2}} c^\dagger_{\bm k j 1 \sigma}c_{\bm k j 2 \sigma} + \right.
\nonumber\\
&&\left. e^{-i \frac{k_z}{2}} c^\dagger_{\bm k j 2 \sigma}c_{\bm k j 1 \sigma} +c.c.\right), 
\nonumber\\
H_{c2}&=&2 t_{c2}  \sum_{\bm k j l \sigma} \cos k_z c^\dagger_{\bm k j l \sigma}c_{\bm k j l \sigma},
\nonumber\\
H_\lambda&=&-2 \lambda \sum_{\bm k j l \sigma} (-1)^l s_\sigma  \left\{ \sin k_1 c^\dagger_{\bm k j l \sigma} c_{\bm k j l \sigma} - \right.
\nonumber\\
&&\left. \sin \frac{k_1}{2} \left(c^\dagger_{\bm k j l \sigma} c_{\bm k j+1 l \sigma}+c.c.\right)\right\}. 
\ee

\item The armchair case
\be
H_t&=& t \sum_{\bm k j l \sigma} \left( 2 \cos \frac{k_1}{2} c^\dagger_{\bm k j l \sigma} c_{\bm k j+1 l \sigma} +  \right.
\nonumber\\
&& \left. c^\dagger_{\bm k j l \sigma} c_{\bm k j+2 l \sigma} + c. c.\right),  
\nonumber\\
H_c&=&2 t_c \sum_{kj\sigma}  \cos \frac{k_z}{2} \times
\nonumber\\
&&\left(e^{i \frac{k_1}{3}} c^\dagger_{\bm k j 2 \sigma} c_{\bm k j 1 \sigma}+e^{i \frac{k_1}{6}} c^\dagger_{\bm k j 1 \sigma} c^\dagger_{\bm k j 2 \sigma} 
+\right.
\nonumber\\
&&\left.e^{-i \frac{k_1}{6}} c^\dagger_{\bm k j 2 \sigma} c^\dagger_{\bm k j 1 \sigma}  + c.c. \right),
\nonumber\\
H_{c2}&=&2 t_{c2}  \sum_{\bm k j l \sigma} \cos k_z c^\dagger_{\bm k j l \sigma}c_{\bm k j l \sigma},
\nonumber\\
H_\lambda&=& - i \lambda \sum_{\bm k j l \sigma} (-1)^l s_\sigma \times
\nonumber\\
&&\left(2 \cos \frac{k_1}{2} c^\dagger_{\bm k j l \sigma} c_{\bm k j+1 l \sigma} - \right.
\nonumber\\
&&\left. c^\dagger_{\bm k j l \sigma} c_{\bm k j+2 l \sigma} \right) 
\nonumber\\
&&+ c. c.
\ee

\end{itemize}

\begin{thebibliography}{99}
\bibitem{chiral-current} 
We should make a remark about the surface charge current. 
It has been pointed out that the surface charge current in a non-$p$-wave chiral 
superconductors is absent in the rotationally symmetric (continuum) system. 
In a lattice system, however, the surface current survives  
when the chiral pairing state is supported by the point group symmetry such as $D_{6h}$ in our system ~\cite{Kalline, Tada-Nie-Oshikawa}.
\end{thebibliography}
\end{document}